\documentclass{article}
\usepackage[utf8]{inputenc}
\usepackage{titling}
\usepackage{lipsum}
\usepackage[space]{grffile}
\usepackage{graphicx}
\usepackage{url}
\usepackage{amsmath}
\usepackage{amssymb}

\usepackage{float}

\title{\huge\bf Sensory Modality Mapping for\\Game Adaptation and Design}
\author{{\large\bf Jeffrey Uhlmann} \\
Dept.\ of Electrical Engineering and Computer Science\\
University of Missouri - Columbia}
\date{}

\begin{document}

\maketitle

\begin{abstract}
In this paper we examine methods for taking game-related information provided in one sensory modality and transforming it to another sensor modality in order to more effectively accommodate sensory-constrained players. We then consider methods for the adaptation and design of games for which gameplay interactions are constrained to a subset of sensory modalities in ways that preserve a common level of novelty-of-experience for players with different sensory capabilities. It is hoped that improved shared experiences can promote interactions among a more diverse spectrum of players.
\end{abstract}

\section{Introduction}

Considerable research has focused on enhancing conventional interactive media technologies to accommodate various types of physical or sensory limitations\footnote{Throughout this paper we use words such as ``limited'' or ``limitations'' to express {\em continuum} constraints on the ability to fully exploit specific information modalities, e.g., partial-vision to blind or hard-of-hearing to deaf, and are not intended to be pejorative in any way.}, and this research has translated into remarkably practical products, e.g., text-to-audio for vision-limited users and audio-to-captions for hearing-limited users.  Physical games (i.e., pre-electronic) were among the earliest to see such accommodations in the form of braille-labeled playing cards and game boards; and many contemporary multimedia games and mixed-reality systems have also been engineered/adapted to accommodate specific types of sensory constraints \cite{fraun,khaliq,neid,ribeiro,rober,yuan}. More recently there has been consideration of games that are natively designed to utilize a subset of sensory modes, e.g., relying entirely on sound or entirely on vision. As we will argue, however, many such games tend to mimic multisensory environments and/or experiences, or analogies thereto, and then {\em by subtraction} reduce the design of gameplay to a subset of sensory modes.  

Consider a conventional first-person video game. Most of the gameplay relies on visual information: sound-effects are primarily cosmetic, i.e., player success would largely be unaffected by their absence, and the remaining nonverbal soundscape (including music) is primarily intended to establish an aesthetically-complementing ambience. The extent to which nonverbal sounds contribute to actual gameplay is predominately limited to the use of stereo panning as a cue for left-right localization of its source. For a person with unconstrained vision, a natural first thought for designing a game that does not rely on vision would be to assume that a realistic aural environment can be created using simple left-right stereo sound panning like that in conventional video games. However, such an assumption verges on condescension. Specifically, the actual aural landscape of the real world includes a rich mix of sound-plus-environment interactions \cite{blauert} from which it is possible to glean a wide variety of contextual cues such as whether one is in a small enclosed space versus a large open space. For example, a virtual forest soundscape (e.g., with crickets and rustling leaves) that remains constant as the player turns and moves could be distracting for a blind person whose expectations tend to be more sensitive than those of a sighted person. In other words, a blind person may find it more difficult to ignore background sounds in a game as ``uninformative'' when in their real-world experience such sounds are their primary source of environmental awareness. For a sighted player, by contrast, visual information in the real world typically assumes primacy to such an extent that natural soundscapes are often referred to as ``background noise.''  

One conclusion that could be drawn is that blind game designers are better suited to develop sound-only games. That is almost certainly true for games that are designed explicitly for blind players, but it is not necessarily true if the goal is to design games that can serve as common shared interests for both sighted and blind people. And of course the same is true in the case of visual-only games for both hearing and hearing-limited people. In this paper we consider both specific and general approaches for transforming sensory information in ways that ``level the playing field'' among persons of differing sensory capabilities by providing a common {\em novelty-of-experience}. 

\section{Sensory Modality Mapping}

We begin by narrowly considering the replacement of a single sensory modality in an existing game. Specifically, consider the need to convey sound information to a player with limited or absent hearing. A common method for doing this is what we will refer to as {\em literal} transformation/translation in the form of annotations (e.g., captions). For example, the sound of a laser blast could be expressed to the player in the form of text saying ``laser blast.'' This is typically easy to implement within a game engine simply by linking the trigger for a particular sound to the display of a particular segment of text. More challenging is the inclusion of relevant contextual information such as ``laser blast to your left,'' which would be conveyed via stereo to a hearing player. However, even if text can be generated that provides fully equivalent information, the experience for a deaf player becomes more descriptive than immersive. This can limit performance in games that rely on reactive interaction because reading not only demands cognitive focus, it also takes time. There is a substantial difference between reflexively reacting to suddenly-heard laser blasts versus reading about the occurrence of laser blasts and deciding how to respond. 

In the case of a limited number of simple textual cues, a deaf player can learn to react to the appearance of a particular caption comparably to the reflexive reaction of a hearing player to the sound. In other words, the visual appearance of the text can become a practically-equivalent proxy for the sound. As video games become increasingly more sophisticated, however, the amount and variety of text necessary to fully summarize the contextual information provided by the overall soundscape will make conscious reading unavoidable for a deaf player. What is needed, therefore, is a translation of the sensory modality -- in this case sound -- to a different sensory modality, e.g., the display of sound information in a purely visual rather than textual form. The specific transformation from sound to visual display must satisfy two requirements:
\begin{enumerate}
\item All of the game-relevant information provided by the sound must be available in the visual representation.
\item Fundamentally orthogonal information components, e.g., simultaneous sounds from independent sources, should be distinguishable.
\end{enumerate}

Simultaneous sounds of laser blasts and moving vehicles are immediately distinguished (segmented) by hearing-able players based on real-world experience with similar categories of sound. What must be emphasized is that the hearing-able person {\em learns} to segment those sounds. S/he would not be able to segment two completely unfamiliar sounds because they would be heard jointly as a single unfamiliar sound until sufficiently-varied exposure experience (learning) makes them distinguishable. What therefore must be provided to the deaf player is a representation that s/he can eventually learn to segment in a way that supports effective gameplay. Ideally, salient information should be apprehended with equal efficiency from either representation by suitably practiced players. In other words, sufficient gameplay experience should permit a deaf player to reflexively respond to the visual representation of sound as effectively as a player who can hear the sound. Of course the deaf player will have an initial learning curve to surmount, but this is not dissimilar to two players introduced to a new video game where one has never before played such a game while the other has vast prior experience with similar games. Sufficient game-specific practice will eventually lead to comparable levels performance with differences that are determined by variables other than level of initial experience.

To make the discussion concrete, consider a monitor display augmented with two narrow vertical display strips on either side as shown in Figure 1, where the audio from stereo-left is transformed to graphical form and displayed on the left vertical strip, and the same is done with respect to the right vertical strip. The physical left-right separation of the two strips immediately conveys whatever spatial information may be connoted from the stereo separation of the soundscape, and what is visually displayed on each strip conveys the remaining information. It is hypothesized that over time a hearing-limited player will become able to discern and segment gameplay-relevant information if the displayed representation of the audio channels satisfies the two previously enumerated requirements.
\begin{figure}
	\label{fig:display}
	\centering
		\includegraphics[scale=1.0]{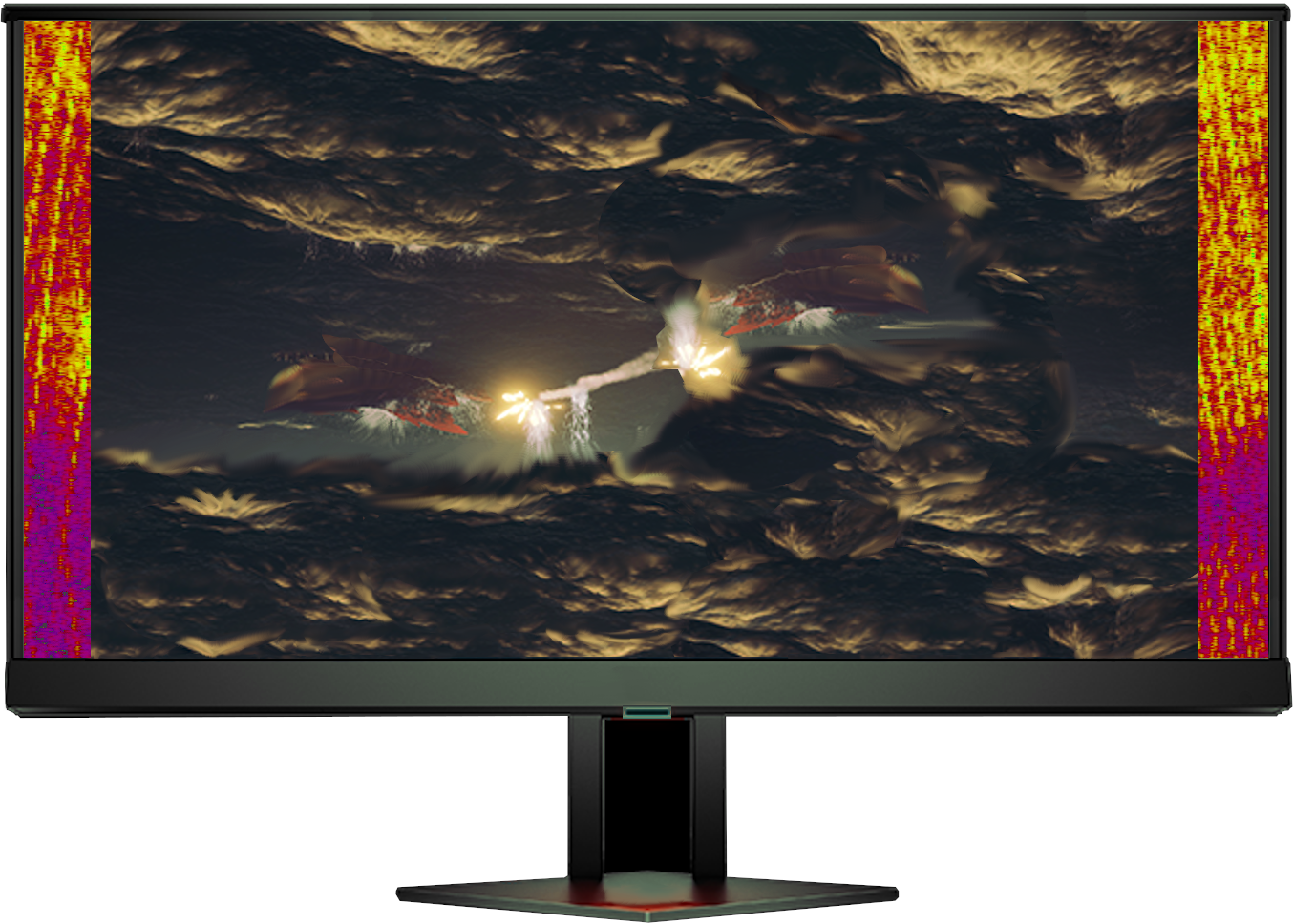}
	\caption{Monitor with left and right vertical strips for displaying audio information that has been transformed to a visual representation. More generally, separate strips could be replaced with a single strip forming a border/frame around the game's video output.}
\end{figure}

This example shows how a single sensory modality can potentially be transformed to a different modality without otherwise affecting gameplay. This is useful for adapting existing video games, and it also provides a glimpse of how game interaction can be more generally transformed so that sensory information is used in a way that is completely different from real-world experience. This will be considered more thoroughly in the next section.

\section{Total Sensory Modality Transformations}

In the previous section we discussed how sound information in a game can be transformed to the visual domain for hearing-limited players. In this section we consider how a game environment can be rendered solely with sound in a manner that has no direct analogy to ordinary experience. This is motivated toward establishing a common novelty-of-experience for both sighted and sight-limied players. We begin by noting that stereo sound is conventionally used to establish horizontal spatial localization of sound sources. Player interpretation of stereo sound in this way is entirely natural because it is directly analogous to how binaural hearing allows sounds to be localized in ordinary real-world experience. In other words, no learning curve is required. For example, we could define a purely sound-based 2-dimensional game environment where stereo information provides horizontal position information while pitch/frequency provides vertical information. In other words, a player can in principle determine the location of an object within a vertical plane (which constitutes the game environment) with stereo left-right panning defining the $x$-axis and low-to-high pitch defining the $y$-axis. A learning curve would be required for players to adapt to this use of pitch to convey spatial information in the vertical direction, but there already exists an intuitive foundation for this because increases in frequency are interpreted linguistically as going from {\em lower} pitches to {\em higher} pitches.

Thinking more generally, any attribute of sound can be used to represent any spatial -- {\em or even non-spatial} -- degree of freedom. For example, pitch could be used to define proximity, where objects that are far away from the player are heard as emitting lower pitches while objects that are close to the player are heard as emitting higher pitches, e.g., a sound that is increasing in pitch represents an object that is moving toward the player. The same could be done using other attributes of sound such as amplitude, timbre, pitch, and waveshape. A particularly natural choice for establishing a horizontally-planar game environment would be to use stereo for the horizontal direction and amplitude for proximity, e.g., nearby objects are louder and distant objects are quieter. However, our goal is to minimize direct analogies to ordinary experience so as to level the playing field in terms of novelty-of-experience for sighted and sight-limited players. Therefore, our preference is to consider completely unfamiliar/novel mappings of sound attributes to spatial dimensions. 

As has been noted, stereo panning is naturally associated with left-right spatial position, and amplitude is associated with proximity. Slightly less connected to ordinary experience is the association of sound frequency with vertical position, as happens when the adjectives {\em lower} and {\em higher} are used to describe relative pitch. If we remove these three from our set of candidate sound attributes, we have timbre and waveshape as variables that have almost no direct association with localization in ordinary experience. Therefore, we can arbitrarily choose to associate them with spatial coordinates to define a 2-dimensional game environment. For example we could associate timbre (``thin/pure'' to ``thick/brassy'') with a $x$-axis and waveshape (``ooh'' to ``aah'') for the $y$-axis to define a Cartesian plane; or we could associate them with range (proximity) and bearing (angle) with respect to player position to define game interactivity in polar coordinates\footnote{Note that an alternative to explicit polar coordinates is to use the familiar {\em ping} of sonar as depicted in countless films and TV episodes involving submarines. More specifically, for each ping heard (and potentially initiated) by the player, a 360-degree sweep is performed over a fixed interval of time with the zero-angle defined as the direction that the player is facing (or, alternatively, directly behind). Thus the relative times between pings from objects reveal their bearings with respect to the player, and timbre (or pitch or other sound attribute) correlates with the range to a pinged object.}.

An issue that has not been discussed thus far is whether player interaction with the game environment is defined in terms of a {\em global} or {\em local} coordinate frame. Global coordinates present difficulties because the ranges of all attribute values must be defined with respect to an arbitrary origin point. This means that as the player moves far away from the origin, the perceived range of attribute values will be at their extreme values and potentially difficult to distinguish. A local coordinate frame resolves such issues because the player's current position and orientation defines the origin, so relative proximity is experienced in a consistent way during gameplay\footnote{Although outside the scope of this paper, the choice to directly provide relative versus global information to the player should affect how game-generated adversaries are simulated. To avoid giving them privileged localization information, bots will need to self-localize based on self-generated local \cite{csorba} or global \cite{julier} maps in a manner analogous to human players.}. In addition, this permits perceived attribute values to be nonlinearly rendered, e.g., so there is greater perceptible resolution for nearby objects at the cost of less discernible resolution for distant objects. Figure 2 shows how a player-centered local Cartesian coordinate frame can be dynamically maintained.

\begin{figure}
	\label{fig:display}
	\centering
		\includegraphics[scale=1.0]{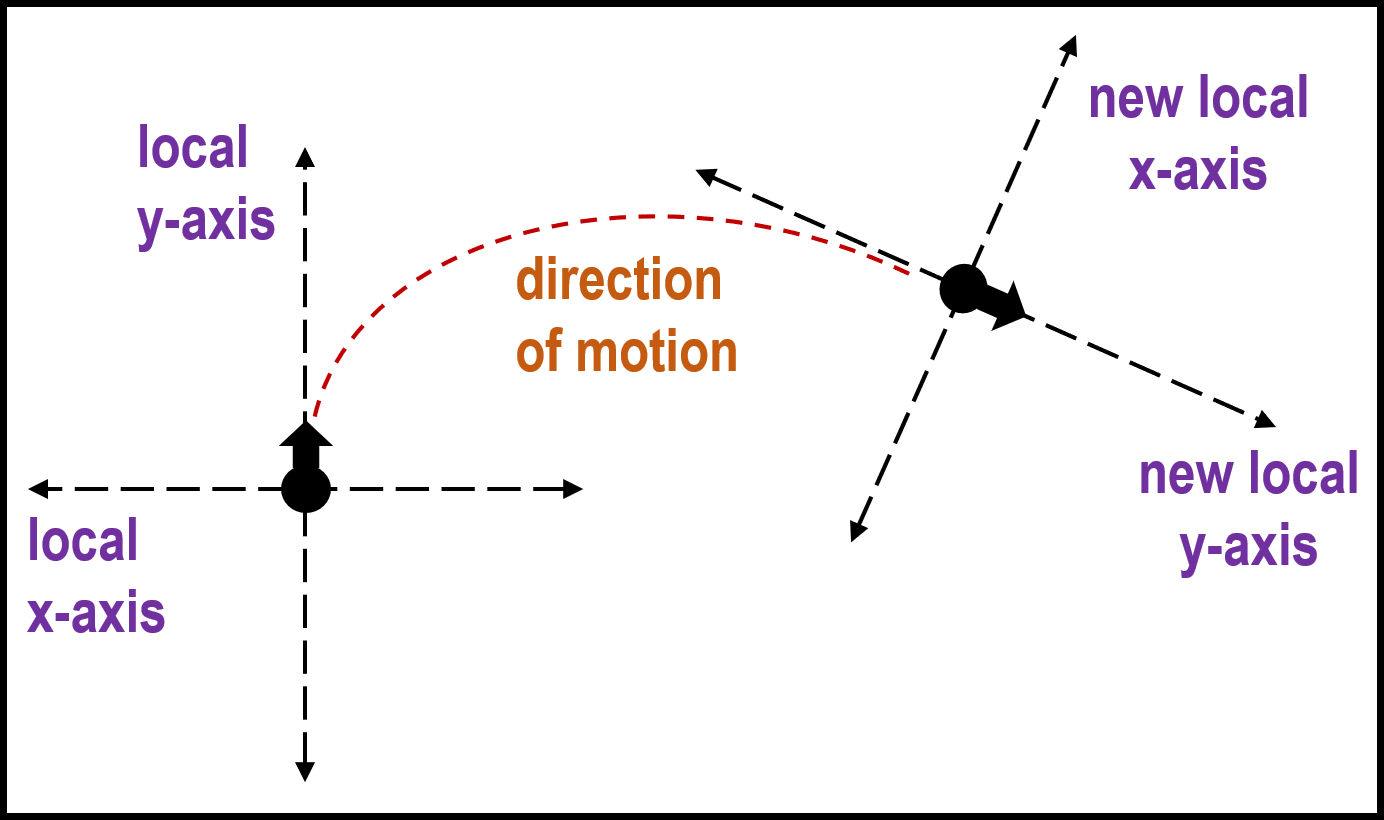}
	\caption{Dynamic local Cartesian coordinate frame centered at player's position with $y$-axis aligned with direction of motion. The player only    receives information about the environment relative to current local coordinate frame.}
\end{figure}

It should be noted that player experience of the environment in the present example can be provided by a single monaural speaker because information relating to each spatial dimension is given by variations in distinct sound attributes, e.g., timbre for $x$ and waveshape for $y$. Thus, any object that is co-located at the position of the player will always be heard with the same spatial sound signature (i.e., timbre-waveshape combination). Similarly, any object that is exactly 10 meters directly in front of the player -- regardless of the player's position in the global coordinate frame -- will also be associated with a distinct sound-attribute combination. Our conjecture is that both sighted and sight-limited players will eventually develop an intuitive cognitive grasp of the spatial game environment, and neither will be initially advantaged nor disadvantaged based on their past experience as sighted or sight-limited. 

\section{A Simple, Concrete Example}

The previous abstract description of the use of sound attributes to convey spatial gameplay information can be conceptually difficult to grasp. In this section we describe an intuitive example that will hopefully clarify and crystalize the main concepts. For this example we will use waveshape to define the $x$-axis and pitch to define the $y$-axis of an implicit Cartesian-planar game environment. The implicit 2-dimensional spatial extent of the game is depicted in Figure 3. 

\begin{figure}
	\label{fig:simpcoord}
	\centering
		\includegraphics[scale=1.0]{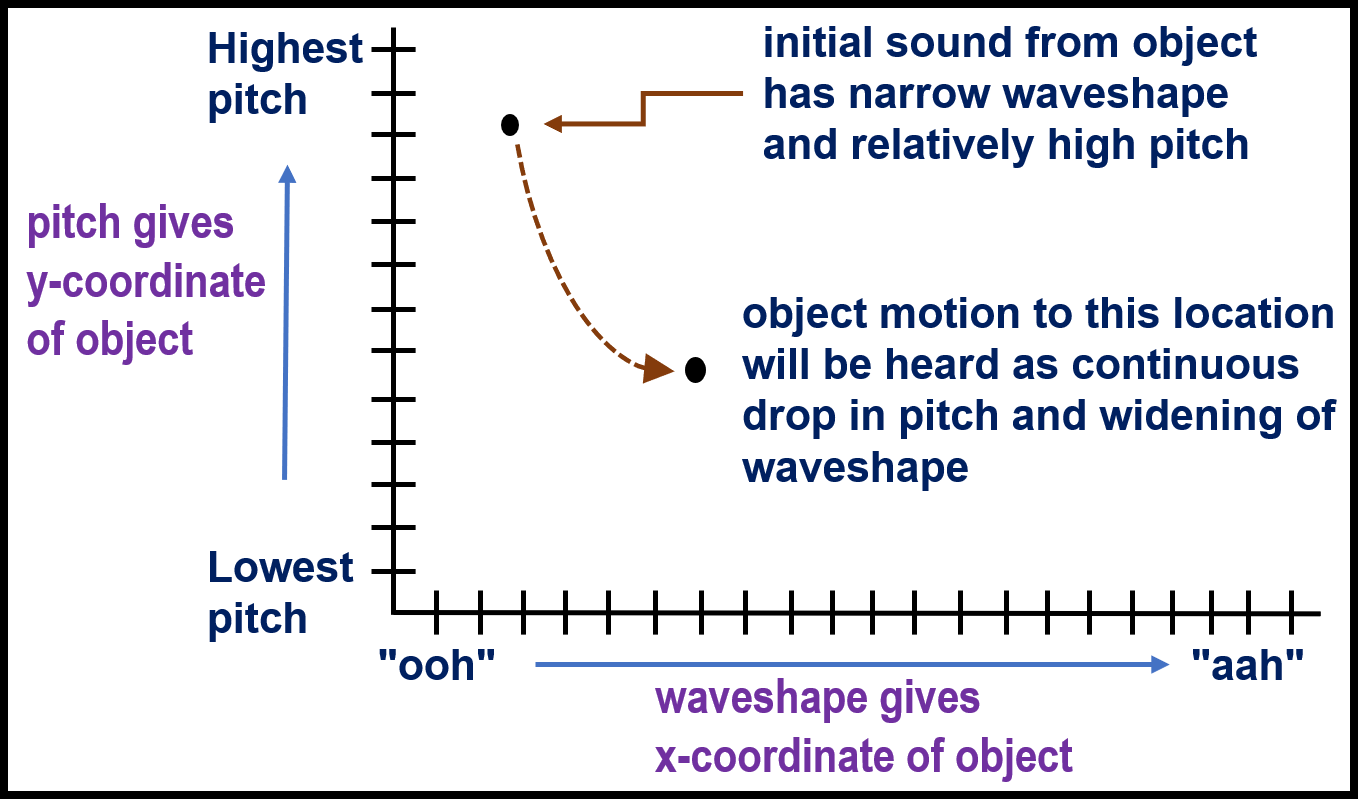}
	\caption{The continuous sound of a moving target in the game is generated based on its changing $x,y$ coordinates, where the waveshape of the sound is correlated with the $x$ value and the pitch is correlated with the $y$ value.}
\end{figure}

In this game we assume a single object, the target, which moves along a randomly-generated continuous path through the game environment. Associated with the target is a continuous sound with waveshape and pitch determined from the target's changing $x,y$ coordinates. As shown in Figure 4, the player is assumed to have a touch-sensitive tablet and will attempt to touch the position on the screen corresponding to the target in order to score points, but touches cannot be registered more frequently than once per second of time. If the target's position is successfully touched then points are awarded and the target resumes from a random location but with faster motion. This means that the player must learn to accurately anticipate future-position from direction-of-motion as the target increases in speed with each iteration. What should be noted from this example is that nothing is displayed on the screen of the tablet because the only source of game information is sound from a single monaural speaker. Thus, from a sensory-capability perspective, a player requires only hearing in one ear and physical ability to manipulate a touch-sensitive tablet.

\begin{figure}
	\label{fig:tablet}
	\centering
		\includegraphics[scale=1.0]{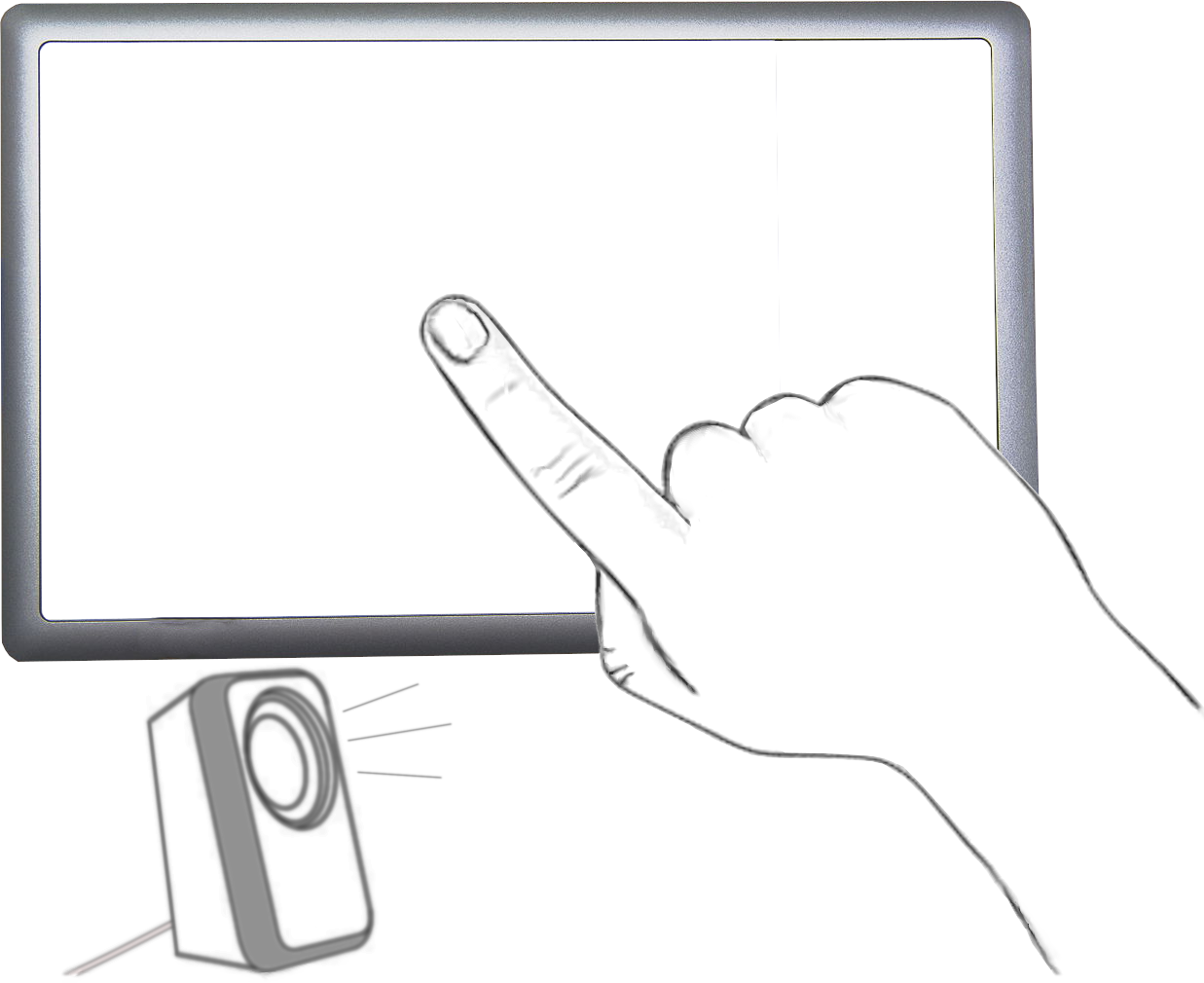}
	\caption{Player presses the screen of a touch-sensitive tablet at the location s/he thinks corresponds to the location of the target based on the combined waveshape and pitch of the sound (see Figure 3). The only source of localization information comes from a monaural speaker, so no graphical information is displayed on the screen.}
\end{figure}

\section{Discussion}

In a previous paper we discussed the use of novel generalizations of mathematical transformations to generalize the operations of simple physical (i.e., non-electronic) games and puzzles for adaptation to exploit the power of computer display devices \cite{uhlmann}. In this paper we have examined the transformation of game output from one sensory modality to another so as to accommodate sensory-limited players. For example, we discussed how output from stereo sound channels could be transformed to visual form to augment the game display for deaf players. We then motivated and examined the transformation of virtual game environments to accommodate sensory-limited players in a manner that neither advantages nor disadvantages anyone with respect to their sensory capability. This was done in support of our expessed goal to motivate future focus on game adaptation and design that emphasizes {\em equal} novelty-of-experience for players across a spectrum of sensory capabilities.

\bibliographystyle{plain}

\end{document}